\def\noi{\noindent}
\def\be{\begin{equation}}
\def\ee{\end{equation}}
\def\ba{\begin{array}}
\def\ea{\end{array}}
\def\bea{\begin{eqnarray}}
\def\eea{\end{eqnarray}}
\def\nn{\nonumber}
\def\um{\frac{1}{2}}
\def\uc{\frac{1}{4}}
\def\tr{{\rm Tr}}
\def\d{{\rm d}}

\documentclass[11pt]{article}

\usepackage[active]{srcltx}
\usepackage{amsfonts}
\usepackage{amsmath}

\begin{document}

\begin{center}
{\Large {\bf Symmetry group of massive Yang-Mills theories\\
without Higgs and their quantization}}
\end{center}

\bigskip
\bigskip

\centerline{V. Aldaya$^{a}$, M. Calixto$^{b,a}$ and F.F. L\'opez-Ruiz$^{a}$}

\bigskip
\centerline{\it $^{a}$ Instituto de Astrof\'\i sica de Andaluc\'{\i}a
(IAA-CSIC),}
\centerline{\it Apartado Postal 3004, Granada 18080, Spain.}
\centerline{\it $^{b}$ Departamento de Matem\'atica Aplicada, Facultad de Ciencias, Universidad de Granada,}
\centerline{\it Campus de Fuentenueva, 18071 Granada, Spain}
\centerline{\it valdaya@iaa.es, calixto@ugr.es, flopez@iaa.es}

\bigskip

\begin{center}

{\bf Abstract}
\end{center}

\small
\setlength{\baselineskip}{12pt}

\begin{list}{}{\setlength{\leftmargin}{3pc}\setlength{\rightmargin}{3pc}}
\item We analyze the symmetry group of massive Yang-Mills theories and their quantization
strongly motivated by an already proposed alternative to the Standard Model of electroweak interactions without Higgs.
In these models the mass generation of the intermediate vector bosons is based on a non-Abelian Stueckelberg mechanism
where the dynamics of the Goldstone-like bosons is addressed by a partial-trace Non-Linear-Sigma
piece of the Lagrangian. In spite of the high non-linearity of the scalar sector, the existence of an infinite
number of symmetries, extending the traditional gauge group, allows us to sketch a group-theoretical quantization
algorithm specially suited to non-linear systems, which departs from usual canonical quantization. On the quantum
representation space of this extended symmetry group, a quantum Hamiltonian preserving the representation can be given,
whose classical analog reproduces the equations of motion.

\end{list}

\normalsize

\noi PACS: 12.15.-y, 12.60.-i, 11.15.-q, 02.20.Tw, 12.10.-g.
\setlength{\baselineskip}{14pt}


\section{Introduction}

In a previous paper \cite{SigmaYM} we proposed an alternative to
the Standard Model of Electroweak Interactions \cite{Weinberg}
without the need for a Spontaneous Symmetry Breaking \`a la
Higgs-Kibble \cite{Higgs}, where the mass of the intermediate
vector bosons was generated by a (gauge invariant) mechanism \`a
la Stueckelberg \cite{StueckelbergRev}, although the kinetic term
corresponding to the scalar bosons parametrizing the gauge group
is only a partial trace on a quotient (coadjoint orbit) $G/H$ of
the gauge group. In fact, $G$  is $SU(2)\times U(1)$ and $H$ is
the ``electromagnetic'' $U(1)$ diagonal subgroup corresponding to
the massless gauge field. There we pointed out that the main
virtue of just considering the partial trace, rather than the sum
over the entire group $G$, in the Lagrangian for the Goldstone
scalars, was the property of possesing an infinite number of
non-gauge symmetries (in the sense of having non-zero associated
Noether invariants) so as to generate the full solution manifold
by means of Noether invariants. The basic symmetry group, actually
a local Euclidean group, was explicitly given. This fact should in
principle provide enough symmetry relationships to face the
quantization of the system according to the standard canonical
quantization algorithm, thus avoiding the apparently unsolvable
dichotomy between unitarity or renormalizability found in the
ordinary (total trace) Non-Abelian Stueckelberg attempts
\cite{dicotomia,Hurth,StueckelbergRev}. However, our intention was
to resort more directly to a proper Group Approach to Quantization
(GAQ for short; see e.g. \cite{23,Ramirez,Sigmita} and references
therein) not subjected to the unavoidable troubles related to
``no-go'' quantization theorems for non-linear systems
\cite{Groenwald,vanHove,Guillemin}.

In this paper we provide the fundamentals of the quantization
formalism that we believe would be the appropriate one to address
a Lagrangian system as the one presented in \cite{SigmaYM}, that
is, of the non-Abelian Stueckelberg type, on the grounds of our
GAQ scheme.

A Group quantization of non-Abelian gauge groups had been only
achieved consistently in $1+1$ dimensions by representing the
corresponding Kac-Moody group \cite{Milikito}. In fact, the
special structure (non-trivial cohomology) of such groups allow
for a central extension \cite{MilikitoII} providing a quantum
representation of the Poisson algebra associated with a WZW-type
Lagrangian. In $3+1$ dimensions, however, the Mickelsson central
extension (two-cocycle) is absent and a bit more involved
construction is required. One of the new required ingredients will
be the consideration of an enlarged symmetry group $G^1(M)$,
containing the gauge group $G(M)$ on the Minkowski spacetime $M$
(the standard gauge group only contribute with null Noether
invariants) parametrized by the Goldstone-like scalar fields
$\phi^a(x)$, along with the corresponding vector potentials
$A_\mu^a(x)$ parametrizing the rest of the new group
\cite{Eduardo,EduardoII}.  The other ingredient refers to the use
of a class of central extensions (two-cocycles) that, even though
they are trivial from some mathematical points of view, they
define  central extensions of the group (and select therefore
specific projective representations of the unextended group)
endowed with a canonical (left- or right-) invariant form
\cite{Julio} which gives a physical Lagrangian for fields living
on a coadjoint orbit of $G^1(M)$. The corresponding Lagrangian can
then be seen as a (covariant) partial-trace of the standard
$\sigma$-model full-trace (chiral) Lagrangian
$Tr(U^{-1}\partial_\mu U U^{-1}\partial^\mu U)$, $U\in G$, coupled
to the vector potentials according to a Minimal Coupling
prescription addressed by the proper structure of the central
extension of the local group. The fact that the Minimal Coupling
in these partial-trace Lagrangians/groups leaves the vector
potentials associated with the subgroup $H$ (characterizing  the
co-adjoint orbit $G/H$) massless, makes the mechanism specially
suited to describe alternatives to the Standard Model without
Higgs particles.

We start in Sec. \ref{classicalsec} by describing  briefly the
Lagrangian formalism corresponding to the non-Abelian Stueckelberg
model on the Minkowski space-time $M$, associated with a gauge
group $G(M)$,  as well as the particular and very interesting case
in which we desire to give mass to only those vector bosons living
in a quotient $G/H$, where $H$ is a compact subgroup. The general
theory is then particularized to the special physical examples
relative to the groups $G=U(1)$, Subsec. \ref{classicalsecU(1)},
and $G=SU(2)$, Subsec. \ref{classicalsecSU(2)}. In Sec.
\ref{quantumsec} we outline the fundamentals of the GAQ and apply
the general group quantization scheme to both examples in SubSecs.
\ref{quantumsecU(1)} and \ref{quantumsecSU(2)}. This results in
the quantum realization of the basic operators representing the
classical Poisson algebra. In the corresponding Hilbert space we
provide a Hamiltonian operator as a quadratic/quartic function of
the basic operators which realize the quantum version of the
corresponding set of classical equations of motion written in
terms of Poisson brackets. Finally, in Sec. \ref{Perturbation}, we
propose a new group-minded perturbative technique which avoids the
eventual troubles associated with the more standard perturbative
techniques when applied to highly non-linear problems. An attempt
to the connection with standard computations is briefly sketched.

\section{The classical theory for the massive Yang-Mills system}\label{classicalsec}

%
The non-Abelian extension of the Stueckelberg formalism
for a general special unitary gauge group $G=SU(n)$ (for a review, see \cite{StueckelbergRev}) consists in giving dynamical content
to the gauge group parameters $\phi^a(x)$ parametrizing a general
local transformation\footnote{To be precise, in the original formalism the scalar
fields $\phi^a(x)$ behave exactly as the local group parameters although they were considered as external
matter fields.} $U(x)=e^{i\phi^a(x)T_a}\in G(M)$, where $T_a, a=1,\dots,{\rm
dim}(G)$ are the Lie-algebra generators of $G$ with commutation
relations $[T_a,T_b]=i C_{ab}^cT_c$. We shall restrict ourselves to unitary groups and set the normalization
$\tr(T_aT_b)=\delta_{ab}$. When referring to the canonical 1-form on $G$, we must distinguish between the left- and
right-invariant ones: $\theta_\mu^L=-iU^\dag\partial_\mu U$ and
 $\theta_\mu\equiv\theta_\mu^R=-i\partial_\mu U U^\dag$,
 respectively ($\theta_\mu=\theta_\mu^aT_a$). The $G$-invariant $\sigma$-model Lagrangian reads:
\be \mathcal{L}_\sigma^{G}=\um \tr(\partial_\mu U \partial^\mu
U^{\dag})=\um\tr(\theta_\mu\theta^\mu)=\um\tr(\theta^L_\mu\theta^{L\mu})\equiv\um
g_{ab}(\phi)\partial_\mu\phi^a\partial^\mu\phi^b\,, \ee
which is highly non-linear and chiral, that is, simultaneously left and right invariant.

The more relevant particularity of sigma-like Lagrangians, with respect to the minimal coupling to  Yang-Mills fields, relies
in the affine or additive way of their coupling to the objects that replace the derivatives of the matter field, namely $\theta_\mu$.
The actual form of this
sort of coupling is a consequence of the specific form of the group action on itself.
In fact, the minimally coupled Lagrangian becomes ($D_\mu\equiv \partial_\mu-A_\mu$)
\be \tilde{\mathcal{L}}_\sigma^{G}=\um \tr((D_\mu U)(D^\mu
U)^{\dag})=\um \tr(
(\theta_\mu-A_\mu)(\theta^\mu-A^\mu)),\label{tlGs}\ee
although $A_\mu$ must be understood as $A_\mu=A_\mu^aT_a$.
Adding the standard kinematical Lagrangian for Yang-Mills fields
\[{\mathcal{L}}_{\rm YM}^G=-\uc \tr(F^{\mu\nu}F_{\mu\nu})\,,\]
where the curvature has the standard expression,

\be F_{\mu\nu}(A)\equiv
\partial_{\mu}A_{\nu}-
\partial_{\nu}A_{\mu}+
[A_{\mu},A_{\nu}]\label{fea},\ee
to the coupled $\sigma$ Lagrangian (\ref{tlGs}), we arrive at the full Lagrangian for Massive
Yang-Mills vector bosons with mass $m$
\be {\mathcal{L}}_{\rm MYM}^{G}={\mathcal{L}}_{\rm
YM}^G+m^2\tilde{\mathcal{L}}_\sigma^{G}\,,\label{chocho}\ee
which is invariant, in particular, under the general gauge transformation $V\in G(M)$
\be U\to VU\,,\;\;\; A_\mu\to VA_\mu V^\dag-i\partial_\mu V
V^\dag.\label{gaugetrans}\ee
As already mentioned in the introduction, this model for massive
Yang-Mills theory, which provides mass to all gauge vector
potentials, cannot be canonically quantized maintaining both
unitarity and renormalizability \cite{dicotomia,Hurth}.

The situation is soundly improved by restricting the whole trace
on $G$ to a partial trace on a quotient manifold $G/H$. $H$ is the
isotropy subgroup of a given direction $\lambda=\lambda^aT_a$, in
the Lie-algebra of $G$, under the adjoint action $\lambda\to
V\lambda V^\dag$, where $\lambda^a$ are real numbers subjected to
$\tr(\lambda^2)=1$. From a strict group-theoretical point of view
and thinking of our specific quantization technique (see later on
Sec. \ref{quantumsec1}), the main advantage of dealing with
partial-trace Lagrangians is that they, or the corresponding
Poincar\'e-Cartan (also named Hilbert or canonical) forms
\cite{RNC,Malliavin,Nair}, can be derived from a
centrally-extended Lie group in much the same way as the
Lagrangian, and the entire (quantum) theory, of a free particle
can be derived from a $U(1)$ central extension of the Galilei
group \cite{23}. And this virtue might be related to the
particular fact that partial-trace Lagrangians can be written as
the square of a total derivative. Actually, by defining
$\Lambda\equiv U\lambda U^\dag$, the claimed $G/H-\sigma$
Lagrangian can be written as:
\be \mathcal{L}_\sigma^{G/H}=\um \tr([-iU^\dag\partial_\mu
U,\lambda]^2)\equiv\um\tr([\theta_\mu^L,\lambda]^2)=\um\tr([\theta_\mu,\Lambda]^2)
=\um\tr((\partial_\mu\Lambda)^2).
\label{trazapar}
\ee
Let us proceed with its minimally coupled version:
\be \tilde{\mathcal{L}}_\sigma^{G/H}=\um \tr([-iU^\dag D_\mu
U,\lambda]^2)=\um\tr([\theta_\mu-A_\mu,\Lambda]^2) \;,\ee
which is again gauge invariant under (\ref{gaugetrans}). As in
(\ref{chocho}), the partial-trace ($G/H$) Massive Yang-Mills
Lagrangian now follows:

\be {\mathcal{L}}_{\rm MYM}^{G/H}={\mathcal{L}}_{\rm
YM}^G+m^2\tilde{\mathcal{L}}_\sigma^{G/H}.\label{ltot1}\ee
We should remark that the change of variables
\be \tilde{A}_\mu=U^\dag(A_\mu-\theta_\mu)U=U^\dag A_\mu U+i
U^\dag\partial_\mu U\;,\label{acurva}\ee
and the fact that $F(A)=U F(\tilde{A}) U^\dag$, renders the
Lagrangian (\ref{ltot1}) into the simple form
\be {\mathcal{L}}_{\rm MYM}^{G/H}=-\uc
\tr(F^{\mu\nu}(\tilde{A})^2)+\um
m^2\tr([\tilde{A}_\mu,\lambda]^2). \label{ltot2}\ee
This change of variables, formally mimicking the shift to the
unitary gauge, turns the actual degrees of freedom of the theory
apparent; that is to say, those of $\hbox{dim}H$ massless vector bosons
and $\hbox{codim}H$ massive ones (see \cite{almeja} for different examples of symmetry ``breaking'' patterns).
In addition, this change of variables must be eventually completed with
the change $\phi=U^\dag\psi$ when the fermionic
matter field $\psi$ will be introduced. It should be remarked that such a redefinition of fermionic
fields has been considered in Literature in order to accomodate mass terms in a way compatible with
gauge invariance (see for instance Ref. \cite{CambioFermi}).

\subsection{The Abelian case}\label{classicalsecU(1)}

The massive Abelian gauge theory clearly corresponds to the simplest version of the total-trace Lagrangian (\ref{chocho}).
In this $G=U(1)$ case, the canonical (left- and right-)invariant 1-form reduces to $\theta_\mu=\partial_\mu\phi$
and the Lagrangian (\ref{chocho}) acquires the simple form:
\be
\mathcal{L}^{U(1)}=-\frac{1}{4}F^{\mu\nu}F_{\mu\nu}+\frac{1}{2}m^2(\partial_\mu\phi-A_\mu)(\partial^\mu\phi-A^\mu).\ee
The Euler-Lagrange equations of motion are:
\be -\partial_\mu F^{\mu\nu}+m^2(\partial^\nu\phi-A^\nu)=0,\;\;
\square\phi=\partial_\nu A^\nu,\ee
where the second one can actually be derived from the first one by
taking the derivative $\partial_\nu$. This is a consequence of the
non-regularity (gauge invariance) of the Lagrangian, which
exhibits only $3$ field degrees of freedom and the corresponding
momenta. In fact, the second equation turns out to be the
transversality equation for the field $\tilde{A}$ (see eq.
(\ref{acurva})), as corresponding to a Proca field.

Introducing the standard ``electric'' and ``magnetic'' notation ($\dot{A}\equiv\partial^0A$):
\be \ba{rcl}E^i\equiv F^{i0}=\partial^iA^0-\dot{A}^i&\rightarrow &
\vec{E}\equiv-\dot{\vec{A}}-\vec\nabla{A^0} \\
B^i\equiv\frac{1}{2}\epsilon^{ijk}F_{jk}&\rightarrow &
\vec{B}\equiv\vec{\nabla}\times\vec{A} \ea\ee
the equations of motions become:
\be\ba{rcl}
\vec{\nabla}\cdot\vec{E}+m^2(A^0-\dot{\phi})&=&0\;\;\;\hbox{Gauss Law}\\
\dot{\vec{E}}-\vec{\nabla}\times\vec{B}-m^2(\vec{A}+\vec{\nabla}\phi)&=&0\;\;\;\hbox{Amp\`ere
Law} \ea\label{GAmpere}\ee
and the Hamiltonian functional reads
\bea H&=&\int d^3x\left[\frac{\partial \mathcal L^{U(1)}}{\partial
\dot A_\nu}\dot A_\nu+\frac{\partial \mathcal L^{U(1)}}{\partial
\dot \phi}\dot\phi -\mathcal L^{U(1)}\right]\nn\\
&=&\int d^3x\left[F^{i0}\dot{A}_i+m^2(\dot{\phi}A^0)\dot{\phi}+\frac{1}{4}(2F_{0i}F^{0i}+F_{jk}F^{jk})\right.\nn\\
&-&\left.\frac{m^2}{2}(\dot{\phi}-A^0)^2-
\frac{m^2}{2}(\partial_i-A_i)(\partial^i-A^i)\right]\nn\\
&=&\int d^3x\left[\frac{1}{2}(\vec{E}^2+\vec{B}^2)+
\frac{m^2}{2}(\dot{\phi}-A^0)^2+\frac{m^2}{2}(\vec{\nabla}\phi+\vec{A})^2-A^0G\right]
\eea
where we have called $G\equiv\vec{\nabla}\cdot\vec{E}-m^2(\dot{\phi}-A^0)$, so that the Gauss law is just $G=0$. This way, the
Hamiltonian density becomes positive on physical trajectories.

\subsection{The Non-Abelian $G=SU(2)$ case}\label{classicalsecSU(2)}

We now make explicit the general classical formulas for the typically non-Abelian case of a partial-trace where the gauge group 
$G$ is $SU(2)$ and the subgroup $H$ is $U(1)$. In that case, some of the
computations are more easily achieved even though the essentials apply to semi-simple compact groups. From formula (\ref{ltot1}),
\be
{\mathcal{L}}_{\rm MYM}^{SU(2)/U(1)}=-\frac{1}{4}F^a_{\mu\nu}F^{\mu\nu}_a+\frac{1}{2}m^2[\theta_\mu-A_\mu,\,\Lambda]^a[\theta^\mu-A^\mu,\,\Lambda]_a\,,
\ee
we derive the momenta,
\bea
\pi_{A^i_a(x)}\equiv\pi^a_i(x)&=&F^a_{i0}=\partial^iA^0_a(x)-\dot{A}^i_a(x)+[A^0, A^i]_a(x)\equiv E^i_a(x)\nn\\
\pi_{\Lambda_a(x)}\equiv\pi^a(x)&=&m^2\dot{\Lambda}^a(x)-m^2[A^0,\Lambda]^a(x)\,,\label{momenta}
\eea
where we have made use of the definition of (non-Abelian) electric field,
to be completed with that of the corresponding magnetic field,
\be
B^i_a(x)\equiv\frac{1}{2}\epsilon^i_{\phantom{i}jk}F^{jk}_a(x)\rightarrow \vec{B}_a\equiv\vec{\nabla}\times\vec{A}_a+
\epsilon_a^{\phantom{a}bc}\vec{A}_b\times\vec{A}_c\,,
\ee
and the equations of motion associated with the variations $\delta A^\nu_a$ and $\delta\Lambda_a$ respectively:
\be
\partial_\mu F^{\mu\nu}_a-[A_\mu, F^{\mu\nu}]_a+m^2\left[[\theta^\nu-A^\nu, \Lambda], \Lambda\right]_a=0\,,\label{eqofmotSYM-PT}
\ee
\be
\partial_\mu[\theta^\mu-A^\mu, \Lambda]-\left[A_\mu, [\theta^\mu-A^\mu, \Lambda], \Lambda\right]=0\,,
\ee
although the latter is recovered from the former by just taking the derivative $\partial_\nu$. For the space components of (\ref{eqofmotSYM-PT}) we  get the Amp\`ere Law,
\be
\dot{E}^i-(\vec{\nabla}\times\vec{B})^i_a-\epsilon^i_{\phantom{i}jk}[A^j, B^k]_a-[A^0, E^i]_a-m^2\left[[\theta^i-A^i, \Lambda],\Lambda\right]_a=0\,,\label{Ampere}
\ee
whereas for the temporal index we obtain the Gauss Law,
\be
\vec{\nabla}\cdot\vec{E}_a-[A_i, E^i]_a+m^2\left[[\theta^0-A^0, \Lambda], \Lambda\right]_a=0\,.\label{Gauss}
\ee

Finally, let us compute the Hamiltonian. From the expression for the momenta (\ref{momenta}) and writing $\dot{A}^i_a$ of $E^i_a$, $A^\mu$ and their spacial derivatives,
we arrive at $ H^{SU(2)/U(1)}_{SYM}$, $H$ for short in the rest of the paper,
\bea
H&\equiv& H^{SU(2)/U(1)}_{SYM}=\int d^3x\left\lbrace E^a_i\dot{A}^i_a+m^2[\theta^0-A^0, \Lambda]_a\dot{\Lambda}^a-{\cal L}^{SU(2)/U(1)}_{SYM}\right\rbrace\nn\\
&=&\frac{1}{2}\int d^3x\left\lbrace\vec{E}^2+\vec{B}^2+m^2[\theta^0-A^0, \Lambda]^2+m^2[\vec{\theta}-\vec{A}, \Lambda]^2-2A^0G^a\right\rbrace\,,
\eea
where we have called, as in the Abelian case,
\be
G\equiv\vec{\nabla}\cdot\vec{E}+[\vec{A}, \vec{E}]+m^2\left[[\theta^0-A^0, \Lambda], \Lambda\right]
\ee
so that the Gauss Law reads simply $G=0$, that which means that the Hamiltonian is again positive on physical trajectories.

\section{Quantization of Massive Yang-Mills Fields}\label{quantumsec}

Linear systems provide a Poisson algebra of classical observables in the solution manifold
(namely, initial position and momenta) realizing the Lie algebra of the Heisenberg-Weyl
group in the corresponding dimension. In field
theories, this suggests to approach the corresponding quantum theory by postulating equal-time
commutation relations between fields and their time derivatives, or conjugate momenta,
$[\phi(\vec{x}), \pi(\vec{y})] = i \delta(\vec{x}-\vec{y})$. In fact, it is worth noticing that these commutators can be
thought of as being an representation of an infinite-dimensional
Lie group. To be precise, and for the simplest physical example of the Klein-Gordon field, we
may write the following group law (see the next subsection for details):
\bea
\phi''(x)&=&\phi'(x)+\phi(x)\nn\\
\phi_\mu''(x)&=&\phi_\mu'(x)+\phi_\mu(x)\nn\\
\zeta''&=&\zeta'\zeta \exp\left\{\frac{i}{2}\int_\Sigma
d\sigma^\mu\left[\phi'(x)\phi_\mu(x)-\phi'_\mu(x)\phi(x)\right]\right\}\,,\label{HW}
\eea
which constitutes a central extension by $U(1)$, parametrized by $\zeta,\, |\zeta|^2=1$, of the Abelian group
parametrized by ``coordinates'' $\phi(x)$ and  (covariant) ``momenta'' $\pi_\mu(x)=\partial_\mu\phi(x)$, and it will be
understood that a quantization of the field $\phi(x)$ is achieved by means of a unitary and irreducible representation
of this group. Fields $\phi(x)$ and $\phi_\mu(x)$ are defined on a Cauchy hyper-surface $\Sigma$ (usually that
characterized by $x^0=0$) or, equivalently, on the entire Minkowski space but, in this case, they must satisfy the
classical equations of motion. It must be clearly stated from now on that $\vec{x}\in\Sigma$ plays the role of an index,
so that the space derivative $\partial_i$ on $\phi$ moves this index. The time component of $\phi_\mu$, however, does
not refer to a (time) derivative of a given $\phi$ and, rather, corresponds to a different family of group parameters,
that is, the momenta (see in this respect Ref. \cite{Vallareport}.

Going to non-linear systems with non-flat phase space would require a more radical departure from the
canonical approach \cite{Isham,Ketov}. This is precisely the situation we are facing now, as a result of
the introduction of the group parameters $\phi$ as physical degrees of
freedom with a (curved) compact target space $G/H$.
We must look for a replacement of the Heisenberg-Weyl
group with a (more involved) symmetry group of the solution
manifold, keeping the general idea of considering as basic
conjugate coordinates those giving central terms under commutation.
As commented in the Introduction, we really aim at a quite
different approach to quantum theory which is totally based on
symmetry grounds and free from the strong limitations which
canonical quantization faces when it tackles highly non-linear
systems. We refer to a Group Approach to Quantization widely
developed in many linear and non-linear examples (see, for
instance \cite{23,Ramirez,Virasoro,Hall}). This formalism
naturally includes the classical theory. The group $G_{MYM}$,
whose unitary and irreducible representations will provide the
quantum theory of our proposed Lagrangian (\ref{ltot1}), will be
given in (\ref{MYMgroup}).

\subsection{Fundamentals of the  general group-quantization formalism}\label{quantumsec1}

The GAQ  formalism is entirely constructed with  canonical structures defined on Lie groups
and the very basic ones consist in the {\it two} mutually {\it commuting}
copies of the Lie algebra $\tilde{\cal G}$ of a group $\tilde{G}$ of {\it
strict symmetry} (of a given physical system), that is, the set of left-
and right-invariant vector fields:

\[{\cal X}^L(\tilde{G})\approx\tilde{\cal G}\approx{\cal
X}^R(\tilde{G})\,,\]
\noi  in such a way that one copy, let us say ${\cal X}^R(\tilde{G})$,
plays the role of {\it pre-Quantum Operators} acting (by usual derivation)
on complex (wave) functions on $\tilde{G}$, whereas the other, ${\cal
X}^L(\tilde{G})$, is used to {\it reduce} the representation in a manner
{\it compatible} with the action of the operators, thus providing the {\it
true quantization}.

In fact, from the group law $g''=g'*g$ of any group $\tilde{G}$, we can
read two different left- and right-actions:
\be g''=g'*g\equiv L_{g'}g,\;\;\;  g''=g'*g\equiv
R_{g}g'.\label{leftrightact}\ee
\noi Both actions commute and so do their respective generators $\tilde{X}^R_a$ and
$\tilde{X}^L_b$, i.e.
$[\tilde{X}^L_a,\;\tilde{X}^R_b]=0\;\;\forall a,b$.


Another manifestation of the commutation between left an right
translations corresponds to the invariance of the left-invariant canonical
1-forms, $\{{\theta^L}^a\}$ (dual to $\{\tilde{X}^L_b\}$, i.e.
${\theta^L}^a(\tilde{X}^L_b)=\delta^a_b$) with respect to the
right-invariant vector fields, that is: $L_{\tilde{X}^R_a}{\theta^L}^b=0$
and the other way around ($L\leftrightarrow R$). In particular, there is
a natural {\it invariant volume} $\omega$ on the group manifold since
we have:
\begin{equation}
L_{\tilde{X}^R_a}({\theta^L}^b\wedge{\theta^L}^c\wedge{\theta^L}^d...)\equiv
L_{\tilde{X}^R_a}\omega=0\,.
\label{omega}
\end{equation}

We should then be able to recover all {\it physical ingredients} of
quantum systems out of {\it algebraic structures}. In particular, the
Poincar\'e-Cartan form $\Theta_{PC}$ and the phase space itself
$M\equiv(x^i,p_j)$ should be regained from a group of {\it strict symmetry
$\tilde{G}$}. In fact, in the special case of a Lie group which bears a
central extension with structure group $U(1)$ parameterized by
$\zeta\in C$ such that $|\zeta|^2=1$, as we are in fact considering, the
group manifold $\tilde{G}$ itself can be endowed with the structure of a
{\it principal bundle} with an {\it invariant connection}, thus
generalizing the notion of {\it quantum manifold}.

More precisely, the $U(1)$-component of the left-invariant canonical form
(dual to the {\it vertical} generator $\tilde{X}^L_\zeta$, i.e.
$\theta^{L(\zeta)}(\tilde{X}^L_\zeta)=1$) will be named {\it quantization
form} $\Theta\equiv{\theta^L}^{(\zeta)}$ and generalizes the
Poincar\'e-Cartan form $\Theta_{PC}$ of Classical Mechanics. The
quantization form remains {\it strictly invariant} under the group
$\tilde{G}$ in the sense that
\[L_{\tilde{X}^R_a}\Theta=0 \ \ \forall a\,,\]

\noi whereas $\Theta_{PC}$ is, in general, only {\it semi-invariant}, that
is to say, it is invariant except for a total differential.

It should be stressed that the construction of a true quantum manifold
in the sense of Geometric Quantization \cite{GQ1,GQ2} can
be achieved by taking in the pair $\langle\tilde{G},\,\Theta\rangle$ the quotient by
the action of the subgroup $G_{\Theta}$ or $G_{\cal C}$, the {\it characteristic subgroup},
generated by those left-invariant vector fields in the kernel of $\Theta$ and
$d\Theta$, that which is called in mathematical terms {\it characteristic
module} of the $1$-form $\Theta$,

\[ {\cal C}_\Theta\;\equiv\;\{\tilde X^L\;/\; i_{\tilde X^L}d\Theta=0=i_{\tilde X^L}\Theta\}.\]
A further quotient by structure subgroup $U(1)$
provides the {\it classical solution Manifold} $M$ or classical {\it phase space}. Even
more, the vector fields in ${\cal C}_\Theta$ constitute the (generalized) {\it classical equations of motion}.

On the other hand, the right-invariant vector fields are used to provide
classical functions on the phase space. In fact, the functions
\[ I_a\;\;\equiv\;\; i_{\tilde{X}^R_a}\Theta\]

\noi are stable under the action of the left-invariant vector fields in
the characteristic module of $\Theta$ (i.e., the generalized equations of motion),
\[ L_{\tilde X^L}I_a=0\;\;\;\;\forall \tilde X^L\in {\cal C}_\Theta\,,\]

\noi and then constitute the {\it Noether invariants}. The Poisson
bracket between two Noether invariants is defined as follows:
\be \{I_a,I_b\}\equiv -
i_{[\tilde{X}^R_a,\tilde{X}^R_b]}\Theta.\label{chochodef} \ee

As a consequence of the central extension structure in $\tilde{G}$, the
Noether invariants (and the corresponding group parameters) are classified
in basic (or dynamical) and non-basic (evolutive or kinematic) depending
on whether or not the corresponding generators produce the central
generator by commutation with some other. Basic Noether
invariants (position and momenta) are paired and independent. Non-basic Noether invariants
(like energy or angular momenta) can be written in terms of the basic ones.

As far as the quantum theory is concerned, the above-mentioned quotient by
the classical equations of motion is really not needed. We consider the
space of complex functions $\Psi$ on the whole group $\tilde{G}$ and
restrict them to only $U(1)$-functions, that is, those which are
homogeneous of degree $1$ on the argument $\zeta\equiv e^{i\phi}\in U(1)$.
Wave functions thus satisfy the $U(1)$-function condition
\be {\tilde{X}^L_\phi\Psi=i\Psi}.\label{u1function}\ee
On these functions the right-invariant vector fields act as {\it
pre-quantum operators} by ordinary derivation. They are, in fact,
Hermitian operators with respect to the scalar product with measure given by the
invariant volume $\omega$ defined above (\ref{omega}).  However, this action is
not a proper quantization of the Poisson algebra of the Noether invariants
(associated with the symplectic structure given by $d\Theta$) since there is a set
of non-trivial operators commuting with this representation. In fact, all the
left-invariant vector fields do commute with the right-invariant ones,
i.e. the pre-quantum operators. According to Schur's Lemma, those operators
must be trivialized in order to achieve full irreducibility. To this end we define a polarization subalgebra as
follows:

\noi A {\it polarization} ${\cal P}$  {\it is a maximal left subalgebra
containing the characteristic subalgebra ${\cal G}_\Theta$ and excluding
the central generator}.

\noi The role of a polarization is that of {\it reducing} the
representation which then constitutes a true {\it quantization}. To this
end we impose on wave functions the (``infinitesimal'') polarization condition:
\[ \tilde{X}^L_b\Psi=0\, \ \ \forall \tilde{X}^L_b\in {\cal P}\,.\]
In finite terms the polarization condition is expressed by the invariance of the wave functions under the finite action
of the Polarization Subgroup $G_P$ acting from the Right, that is:
\be
\Psi(g'g_P)=\Psi(g')\ \ \ \forall g_P\in G_P\,. \label{PolarizacionF}
\ee

To be intuitive, a polarization is made of half the left-invariant vector
fields associated with basic (independent) variables of the solution
manifold (either positions or momenta in the simplest case of the free Galilean particle)
in addition to those associated with kinematic parameters as time
or rotational angles. We should remark that the classification
above-mentioned of the Noether invariants in basic and non-basic also
applies to the quantum operators so that the latter ones are written in
terms of the formers.

\subsection{The Massive Electromagnetic Quantization Group}\label{quantumsecU(1)}

In this subsection we attempt to the quantization of the standard ($U(1)$) Stueckelberg system which classical
theory has been sketched in subsection \ref{classicalsecU(1)}. This will constitute an introductory example to
the more intrincate case of the Non-Abelian theory. To this end, let us write a basic group
law appropriate for the general GAQ formalism to account for both the classical and quantum
theory. We have:
\bea
\varphi''(x)&=&\varphi'(x)+\varphi(x)\nn\\
\varphi_\mu''(x)n^\mu&=&\varphi_\mu'(x)n^\mu+\varphi_\mu(x)n^\mu\nn\\
{\cal {\cal A}}_\mu''(x)&=&{\cal A}_\mu'(x)+{\cal A}_\mu(x)  \label{EMgroup}\\
{\cal F}_{\mu\nu}''(x)&=&{\cal F}_{\mu\nu}'(x)+{\cal F}_{\mu\nu}(x)\nn\\
\zeta''&=&\zeta'\zeta \exp\left(\frac{i}{2}\int_\Sigma
d\sigma^\mu(x)J_\mu(\varphi',{\cal A}',{\cal F}';\varphi,{\cal A},{\cal F})\right),\nn\\
J_\mu&=&J_\mu^{\cal A}+J_\mu^\varphi,\nn\\
 J_\mu^{\cal A}&=&({\cal A}'^\nu(x)-\partial^\nu\varphi'(x)){\cal F}_{\mu\nu}(x)-{\cal F}'_{\mu\nu}(x)({\cal A}^\nu(x)-\partial^\nu\varphi(x))\nn\\
J_\mu^\varphi&=&m^2\left(\varphi'(x)(\partial_\mu\varphi(x)-{\cal
A}_\mu(x))- (\partial_\mu\varphi'(x)-{\cal
A}'_\mu(x))\varphi(x)\right) \,.\nn \eea

As commented in the case of the Klein-Gordon group (\ref{HW}), the fields in this group law are supposed to be defined on
a Cauchy hypersurface $\Sigma$ (characterized by an unit vector $n^\mu$, usually $n=(1,0,0,0)$) or, otherwise, if defined on the entire Minkowski space-time, $M$, they should
satisfy the equations of motion, a fact which would be reflected by the conservation of the current density $J_\mu$ defining
the (co-cycle of) the central extension, that is, $\partial^\mu J_\mu=0$ \footnote{The conservation of the current, when the field
parameters were defined on the entire $M$, is achieved if the fields would verify $\partial^\mu{\cal F}_{\mu\nu}+m^2{\cal A}_\nu=0$
and $\partial^\mu\partial_\mu\varphi-\partial^\mu{\cal A}_\mu=0$, which correspond to the Euler-Lagrange equations after the
identification ${\cal A}_\mu\equiv A_\mu-\partial_\mu\phi$ and $\varphi\equiv\phi+\frac{1}{m^2}\partial^\nu A_\nu$.}. In
Ref. \cite{Miguel} the group law relative to the pure electromagnetic field, with field parameters defined on $M$, and including
the Poincar\'e subgroup, was considered. Here, this subgroup is discarded on behalf of simplicity, since we are primarily interested
in those symmetries providing the basic Noether Invariants parametrising the classical solution manifold. However, the time evolution
is eventually recovered by providing a Hamiltonian $\mathbb{H}$ as a function of the basic Noether Invariants. In fact, the Poisson bracket with $\mathbb{H}$,
will recover the classical equations of motion.

Let us compute from the group law the right- and left-invariant vector fields, the Lie algebra commutators, quantization form and
Noether Invariants. By deriving the group composition parameters $g''$ with respect to each one of $g'$ at the identity $g'=\hbox{Id}$ we obtain:
\bea
\tilde{X}^R_{\varphi(x)}&=&\frac{\delta}{\delta\varphi(x)}+\frac{1}{2}\left(\partial_i{\cal F}^{0i}+m^2(\partial_\mu\varphi-{\cal A}_\mu) n^\mu\right)\Xi\nn\\
\tilde{X}^R_{\varphi^\mu(x)}&=&\frac{\delta}{\delta\varphi^\mu(x)}-\frac{1}{2}m^2\varphi(x)n_\mu\Xi\nn\\
\tilde{X}^R_{{\cal A}^\mu(x)}&=&\frac{\delta}{\delta{\cal A}^\mu(x)}+\frac{1}{2}\left({\cal F}_{\nu\mu}(x)+m^2\varphi(x)
\eta_{\mu\nu}\right)n^\nu\Xi\label{RightAbelian}\\
\tilde{X}^R_{{\cal F}^{\mu\nu}(x)}&=&\frac{\delta}{\delta{\cal F}^{\mu\nu}(x)}+\frac{1}{2}\left((\partial_\nu\varphi(x)-{\cal A}_\nu(x))n_\mu-(\partial_\mu\varphi(x)-
       {\cal A}_\mu(x))n_\nu\right)\Xi\,.\nn
\eea
The non-zero Lie brackets among them are:
\bea
\left[\tilde{X}^R_{\varphi(x)},\,\tilde{X}^R_{\varphi^\mu(y)}\right]&=&-m^2n_\mu\delta(x-y)\Xi\nn\\
\left[\tilde{X}^R_{\varphi(x)},\,\tilde{X}^R_{{\cal A}^\mu(y)}\right]&=&m^2n_\mu\delta(x-y)\Xi\nn\\
\left[\tilde{X}^R_{\varphi(x)},\,\tilde{X}^R_{{\cal F}^{\mu\nu}(y)}\right]&=&-n_{[\nu}\partial_{\mu]}^x\delta(x-y)\Xi\nn\\
\left[\tilde{X}^R_{{\cal A}^\mu(x)},\,\tilde{X}^R_{{\cal F}^{\sigma\nu}(y)}\right]&=&-\eta_{\mu[\nu}n_{\sigma]}\delta(x-y)\Xi\label{conmutatas}
\eea

 In the same way, turning primed parameters to non-primed ones we get:
\bea
\tilde{X}^L_{\varphi(x)}&=&\frac{\delta}{\delta\varphi(x)}-\frac{1}{2}\left(\partial_i{\cal F}^{0i}(x)+
     m^2(\partial_\mu\varphi(x)-{\cal A}_\mu(x)) n^\mu\right)\Xi\nn\\
\tilde{X}^L_{\varphi^\mu(x)}&=&\frac{\delta}{\delta\varphi^\mu(x)}+\frac{1}{2}m^2\varphi(x)n_\mu\Xi\nn\\
\tilde{X}^L_{{\cal A}^\mu(x)}&=&\frac{\delta}{\delta{\cal A}^\mu(x)}-\frac{1}{2}\left({\cal F}_{\nu\mu}(x)-m^2\varphi(x)
\eta_{\mu\nu}\right)n^\nu\Xi\label{LeftAbelian}\\
\tilde{X}^L_{{\cal F}^{\mu\nu}(x)}&=&\frac{\delta}{\delta{\cal F}^{\mu\nu}(x)}-\frac{1}{2}\left((\partial_\nu\varphi(x)-{\cal A}_\nu(x))n_\mu-(\partial_\mu\varphi(x)-
       {\cal A}_\mu(x))n_\nu\right)\Xi\,.\nn
\eea

By duality on the left-invariant generators (\ref{LeftAbelian})  we compute the quantization $1$-form $\Theta ^{U(1)}$:
\bea
\Theta^{U(1)}&=&\frac{1}{2}\int_\Sigma d\sigma^\mu\left({\cal F}_{\mu\nu}\delta({\cal A}^\nu-\partial^\nu\varphi)-
      ({\cal A}^\nu-\partial^\nu\varphi)\delta{\cal F}_{\mu\nu}\right.\nn\\
&+&\left. m^2[\varphi\delta({\cal A}_\mu-\partial_\mu\varphi)-({\cal A}_\mu-\partial_\mu\varphi)\delta\varphi]\right)+\frac{d\zeta}{i\zeta}\,.\nn
\eea
and from it the Noether Invariants $I=i_{\tilde{X}^R}\Theta^{U(1)}$ (we choose for $\Sigma$, $x^0=0$). They are:
\bea
I_{\varphi(x)}&=&m^2(\dot{\varphi}-{\cal A}^0)(x)\equiv -m^2\mathbb{A}^0(x)-\vec{\nabla}\cdot\vec{\mathbb{E}}(x)\equiv -\mathbb{G}\nn\\
I_{\dot{\varphi}(x)}&=&-m^2\varphi(x)\nn\\
I_{{\cal A}^0(x)}&=&m^2\varphi(x)\equiv-\mathbb{E}_0\nn\\
I_{{\cal A}^i(x)}&=&-{\cal F}_{i0}(x)\equiv-\mathbb{E}_i\label{NoetherU(1)}\\
I_{{\cal F}^{i0}(x)}&=&{\cal A}_i(x)-\partial_i\varphi(x)\equiv\mathbb{A}_i(x)\nn
\eea
where we observe the identity $I_{\dot{\varphi}(x)}+I_{{\cal A}^0(x)}=0$ which expresses that the associated generator
is gauge, that is, the combination $\tilde{X}^R_{\dot{\varphi(x)}}+\tilde{X}^R_{{\cal A}^0(x)}$ has a null Noether Invariant.

According to the general definition of Poisson bracket in our group-theoretical approach, expression (\ref{chochodef}),  we arrive at the
Poisson algebra (non-zero brackets only):
\bea
\lbrace\mathbb{E}_j(x),\,\mathbb{A}_k(y)\rbrace&=&\eta_{jk}\delta(x-y)\nn\\
\lbrace\mathbb{E}_0(x),\,\mathbb{A}_0(y)\rbrace&=&-\delta(x-y)\nn\\
\lbrace\mathbb{G}(x),\,\mathbb{E}^0(y)\rbrace&=&m^2\delta(x-y)\label{chochoP}\\
\lbrace\mathbb{G}(x),\,\mathbb{A}_i(y)\rbrace&=&\partial^x_i\delta(x-y)\nn
\eea

Time evolution can be recovered from the solution manifold by
writing down classical equations of motion \`a la Hamilton
associated with a Hamiltonian function of the Noether Invariants
which reproduces physical trajectories in time. In fact, the
Hamiltonian
\be \mathbb{H}=\frac{1}{2}\int d^3y\left(\vec{\mathbb{E}}^2+
\vec{\mathbb{B}}^2+m^2\mathbb{A}_0^2+m^2(\vec{\mathbb{A}}+
\frac{\vec{\nabla}\mathbb{E}_0}{m^2})^2-2\mathbb{A}_0\mathbb{G}\right)
\label{AmilAbel} \ee
leads under the Poisson bracket (\ref{chochoP}) with $\mathbb{E}$
and $\mathbb{A}$ to:
\bea
\dot{\vec{\mathbb{E}}}(x)&=&\lbrace\vec{\mathbb{E}}(x),\,\mathbb{H}\rbrace=
\vec{\nabla}\times\vec{\mathbb{B}}(x)+
    m^2\left(\vec{\mathbb{A}}(x)+\frac{\vec{\nabla}\mathbb{E}_0(x)}{m^2}\right)\nn\\
\dot{\vec{\mathbb{A}}}(x)&=&\lbrace\vec{\mathbb{A}},\,\mathbb{H}\rbrace=-\vec{\mathbb{E}}(x)-\vec{\nabla}\mathbb{A}^0(x)\nn\\
\dot{\mathbb{E}}^0(x)&=&\lbrace\mathbb{E}^0(x),\,\mathbb{H}\rbrace=\mathbb{G}(x)=\vec{\nabla}\cdot\vec{\mathbb{E}}(x)+m^2\mathbb{A}^0(x)\nn\\
\dot{\mathbb{G}}(x)&=&\lbrace\mathbb{G}(x),\,\mathbb{H}\rbrace=0\,,\nn
\eea
the last equation simply expressing the gauge invariance of
$\mathbb{H}$. They reproduce the Lagrangian equations of motion
(\ref{GAmpere}) under the identification $\mathbb{A}^i\leftrightarrow A^i, \mathbb{A}^0 \leftrightarrow A^0-\dot\phi, \, 
\vec{\mathbb{E}}\leftrightarrow\vec{E}$ and $\mathbb{E}_0\leftrightarrow m^2\phi$.

\subsubsection{The Quantum Representation}\label{quantumsubsecU(1)}

According to the general scheme of GAQ we start from complex
$U(1)$-functions $\Psi$ on the entire centrally-extended group to
be represented, that is, functions which are homogeneous of degree
one on the parameter $\zeta\in U(1)$. In order to obtain an
irreducible representation these functions must be restricted by
the Polarization condition established by means of a polarization
subgroup $G_P$ of the finite left action. A look at the Lie
algebra commutators of our symmetry group reveals that the
characteristic subgroup $G_{\cal C}$ contains elements of the
local $U(1)$ subgroup as well as elements of the tangent group for
which ${\cal A}$ equals $\partial_\mu\varphi$, that is,
\be g_{{\cal C}}=(\varphi=0,\varphi_\mu n^\mu,{\cal
A}_\mu=\varphi_\mu,{\cal F}_{\nu\sigma}=0,\zeta=1)\,, \ee
and the  Polarization subgroup (see (\ref{PolarizacionF}))
contains, in addition, the tangent elements with arbitrary values
of $\check{{\cal A}}\equiv({\cal A}-\partial\varphi)$. We then
have:
\be g_P=(\varphi=0,\varphi_\mu n^\mu, {\cal A}_\mu,{\cal
F}_{\nu\sigma}=0,\zeta=1)\,. \ee
The polarization subgroup acts on the original complex functions
$\Psi(g')=\Psi(\varphi',\partial_\mu\varphi',{\cal A}'_\nu, {\cal
F}'_{\mu\nu},\zeta')$ from the right:
$\Psi(g')\rightarrow\Psi(g'g_P)$. The key point in searching for
the appropriate form of the wave functions, invariant under the
Polarization subgroup (\ref{PolarizacionF}), is to notice the
factor which appears in front of the wave function as a
consequence of the co-cycle in the composition law of the central
$U(1)$ argument ($\Psi$ is homogeneous of degree one on it). This
factor must be cancelled out by some $U(1)$ term factorizing an
arbitrary function $\Phi$ of given arguments. It is easy to see
that the following wave function $\psi$ is the general solution to
the polarization equation (\ref{PolarizacionF}):
\be \Psi(g)=\zeta e^{\frac{i}{2}\int_\Sigma d\sigma^\mu\left(m^2
\check{\cal A}_\mu\varphi+
      \check{\cal A}^\nu {\cal F}_{\mu\nu}\right)}
      \Phi(\varphi,{\cal F}_{\mu\nu} n^\nu)\label{wavefunction(u1)}
\ee
where $\Phi$ is an arbitrary function of its arguments. In fact,
choosing $d\sigma^\mu$ as $d^3x$, without loss of generality,
\bea \Psi(g'g_P)&=&\zeta'e^{\frac{i}{2}\int d^3x\left(-{\cal
F}'_{0i}{\cal A}^i+m^2\varphi'(\varphi_0-{\cal A}_0)\right)}
e^{\frac{i}{2}\int d^3x\left((\check{\cal A}'^i+\check{\cal
A}^i){\cal F}'_{0i}+m^2(\check{\cal A}'_0+\check{\cal
A}_0)\varphi'\right)}\Phi(\varphi',{\cal F}'_{i0}) \nn\\
&=& \zeta' e^{\frac{i}{2}\int d^3x\left(\check{\cal A}'^i{\cal
F}'_{0i}+m^2\check{\cal A}'_0\varphi'\right)}\Phi(\varphi',{\cal
F}'_{i0}) =\Psi(g') \eea
%

%
For the usual choice of $\Sigma$, the arbitrary part $\Phi$ of
$\Psi$ can be written in terms of the Noether invariants
(\ref{NoetherU(1)}). That is, $\Phi=\Phi(\mathbb{E}_\mu)$, thus
arriving at the ``electric field representation''.

The action of the right-invariant vector fields preserve the space
of polarized wave functions, due to the commutativity of the left and right actions
as  already stated, so that it is possible to define an action of them
on the arbitrary factor $\Phi$ in the wave functions. It is not difficult to
demonstrate that on this space of functions the quantum operators
acquire the following expression:
\begin{align}
\hat{E}^\mu\Phi&\equiv\phantom{-} i \zeta^{-1}
e^{-\frac{i}{2}\int_\Sigma d\sigma^\mu \left(m^2\check{{\cal
A}}_\mu\varphi+\check{{\cal A}}^\nu {\cal
F}_{\mu\nu}\right)}\tilde{X}^R_{{\cal A}_\mu}\Psi
=\mathbb{E}^\mu\Phi\nn\\
\hat{A}^i\Phi&\equiv -i \zeta^{-1}e^{-\frac{i}{2}\int_\Sigma
d\sigma^\mu \left(m^2\check{{\cal A}}_\mu\varphi+\check{{\cal
A}}^\nu {\cal F}_{\mu\nu}\right)}\tilde{X}^R_{{\cal F}_{0i}}\Psi
=-i\frac{\!\!\delta}{\delta \mathbb{E}^i}\Phi\label{operatas(u1)}\\
\hat{G}\Phi&\equiv -i \zeta^{-1} e^{-\frac{i}{2}\int_\Sigma
d\sigma^\mu \left(m^2\check{{\cal A}}_\mu\varphi+\check{{\cal
A}}^\nu {\cal F}_{\mu\nu}\right)}\tilde{X}^R_{\varphi}\Psi
=\left(\vec{\nabla}\cdot\vec{\mathbb{E}} +
m^2\frac{\!\!\delta}{\delta \mathbb{E}_0}\right)\Phi\,.\nn\\
\hat{A}_0\Phi&\equiv
\frac{1}{m^2}(\hat{G}-\vec{\nabla}\cdot\hat{\vec{\mathbb{E}}})\Phi\nn
\end{align}
The constraint condition $\hat{G}\Phi=0$, if required, would
account for the quantum implementation of the ``Gauss law".

On the quantum representation space we can construct the
Hamiltonian operator $\hat{H}$ that represents, without ambiguity,
the classical Hamiltonian (\ref{AmilAbel}):
\be \hat{H}\Phi=\frac{1}{2}\int
\d^3x\left(\hat{\vec{E}}^2+\hat{\vec{B}}^2+m^2\hat{A}_0^2
+m^2(\hat{\vec{A}}+\frac{\vec{\nabla}\hat{E}_0}{m^2})^2-2\hat{A}_0\hat{G}\right)\Phi\,.
\ee
It is again quite relevant the fact that this operator preserves the Hilbert space of quantum states.

\subsection{The Massive Yang-Mills Quantization Group}\label{quantumsecSU(2)}

Now that the simpler Abelian case has been reviewed under the
general GAQ scheme, we are in conditions to tackle the more
involved non-Abelian massive Yang-Mills theory under the same
framework. To this end, we shall write down a group law,
$G_{MYM}$, providing both the classical and quantum description.
That is,
\bea
U''(x)&=&U'(x)U(x) \nn\\
\vartheta''_\mu(x)n^\mu&=& U'(x)\vartheta_\mu(x)n^\mu U'^\dag(x)+\vartheta'_\mu(x)n^\mu,\nn\\
{\cal A}''_\mu(x)&=& U'(x){\cal A}_\mu(x) U'^\dag(x)+{\cal A}'_\mu(x),\nn\\
{\cal F}''_{\mu\nu}(x)&=& U'(x){\cal F}_{\mu\nu}(x) U'^\dag(x)+{\cal F}'_{\mu\nu}(x), \label{MYMgroup}\\ 
\zeta''&=&\zeta'\zeta \exp\left(i\int_\Sigma
d\sigma^\mu(x)J_\mu(U',{\cal A}',{\cal F}';U,{\cal A},{\cal F})\right),\nn\\
J_\mu&=&J_\mu^{{\rm YM}}+J_\mu^\sigma,\nn\\
 J_\mu^{{\rm YM}}&=&\tr\left(
 ({\cal A}'^\nu-\vartheta'^\nu)U'{\cal F}_{\mu\nu}U'^\dag-{\cal F}'_{\mu\nu}U'({\cal A}^\nu-
 \vartheta^\nu)U'^\dag\right),\nn\\
J_\mu^\sigma&=&m^2 \tr\left(\lambda(U'({\cal
A}_\mu-\vartheta_\mu)U'^\dag-({\cal
A}_\mu-\vartheta_\mu))\right)\,,\nn \eea
where all fields are assumed to be defined on the Cauchy surface
$\Sigma$, so that, the time translation can not be directly
implemented, in  contrast with the case of free fields
\cite{Vallareport}. However, as in the Abelian case, we shall
construct an explicit Hamiltonian operator to account for the time
evolution on the quantum states (see below)\footnote{The
coboundary piece $J^\sigma_\mu$ in (\ref{MYMgroup}) does not
exhaust all possibilities of encoding dynamical content in the
group. In fact, the coboundary current ${J'_\mu}^\sigma=i\kappa
n_\mu\tr\left(\lambda(\ln(U'U)-\ln(U')-\ln(U))\right)$ was
considered in \cite{ymass}, as a preliminary attempt to give mass
to vector bosons, but we shall not pursue this possibility any
further here.}. This group constitutes the minimal symmetry
necessary to reproduce the solution manifold associated with the
classical Lagrangian (\ref{ltot1}). It is also worth mentioning
that the parameter ${\cal F}$ appearing in the group law must be a
gauge covariant function of ${\cal A}$ and $\partial{\cal A}$, and
the simplest one is well-known to be the same function as in
(\ref{fea}).

Before proceeding further in any actual computation according to
the general GAQ scheme, let us insist on several facts. Firstly,
the objects ${\cal A}^a_\mu$ and $\vartheta^a_\mu$ behave exactly
in the same manner under the local group $G(M)$ and
$\vartheta^a_i$ (spacial components) are particular cases of
${\cal A}^a_i$, that is, the cases in which the Yang-Mills
potentials are pure gauge. Secondly, there are some extra
unexpected parameters ${\cal A}^a_0$, with respect to the massless
case, which are the zero (time) components of those vector
potentials living on the orbit $G/H$. They also behave as the time
components of $\vartheta^a_\mu$ but they never are pure gauge, as
they fix the initial values of the derivatives of the fields
$\varphi^a$ on the Cauchy surface $\Sigma$. We then require
non-trivial (non-gauge) four components ${\cal A}^a_\mu$ for each
index $a$ of the orbit of $G$. We accordingly add a time component
for the ``electric'' field ${\cal E}^a$, ${\cal E}^a_i={\cal
F}^a_{i0}$, namely ${\cal E}^a_0=m^2\tr(T^a\Lambda)$,
constituting, somehow, a new component, ``${\cal F}_{00}$'', of
the curvature ${\cal F}^a_{\mu\nu}$. From the mathematical point
of view, the appearance of this non-conventional field degrees of
freedom are a consequence of the piece $J^\sigma$ of the current
in (\ref{MYMgroup}).


From this group law (\ref{MYMgroup}), the generators of the left group  action (the right-invariant vector fields)
can be written:
\bea \tilde X^{R}_{\varphi^a (x)} &=& X^{R\;(G)}_{\varphi^a(x)}
-C_{ab}^c\left(\vartheta^b_\nu(x)n^\nu\frac{\delta\;\;}{\delta\vartheta^c_\mu(x)n^\mu}
+ {\cal A}^b_\nu(x) \frac{\delta\;\;}{\delta {\cal A}^c_\nu(x)}
-C_{ab}^c{\cal F}^b_{\mu\nu}(x)\frac{\delta\;\;}{\delta {\cal F}^c_{\mu\nu}(x)}
  \right)\nn\\
&-&\left(C_{ab}^c m ({\cal A}^b_\nu(x)-\vartheta^b_\nu(x))\lambda_c
+\partial^\mu {\cal F}_{a\mu\nu}\right)n^\nu(x)\Xi\nn\\
\tilde X^{R}_{\vartheta^a_{\mu}(x)n^\mu} &=&\frac{\delta\;\;}{\delta \vartheta^a_{\mu}(x)n^\mu} \nn\\
\tilde X^{R}_{{\cal A}^a_{\mu}(x)} &=&\frac{\delta\;\;}{\delta
{\cal A}^a_{\mu}(x)}-
{\cal F}^{\mu\nu}_a(x)n_\nu(x)\Xi\nn\\
\tilde X^{R}_{{\cal F}^a_{\mu\nu}(x)} &=&\frac{\delta\;\;}{\delta {\cal F}^a_{\mu\nu}(x)}
    -\frac{1}{2}\left(({\cal A}^\mu_a(x)-\vartheta^\mu_a(x))n^\nu(x)-({\cal A}^\nu_a(x)-\vartheta^\nu_a(x))n^\mu\right) \Xi\nn \\
\tilde X^{R}_{\zeta} &=& \hbox{Re}(i\zeta
\frac{\partial}{\partial\zeta})\equiv \Xi\,.\label{ym-XR} \eea
The corresponding non-null (equal-time) Lie bracket are:
\bea
\Bigl[\Bigl[\tilde{X}^R_{\varphi^a (x)}, \tilde{X}^R_{\varphi^b(y)}\Bigr] \Bigr]&=& -C_{ab}^c \delta(x-y)
\tilde{X}^R_{\varphi^c(x)}\nn\\
\Bigl[\Bigl[\tilde{X}^R_{\varphi^a(x)},\tilde{X}^R_{\vartheta^b_\mu(y)n^\mu}\Bigr] \Bigr] &=& -C_{ab}^c \delta(x-y)\tilde{X}^R_{\vartheta^c_\mu(x)n^\mu}
+m^2C_{ab}^c\lambda_c\delta^\mu_0\delta(x-y)\Xi\nn\\
\Bigl[\Bigl[\tilde{X}^R_{\varphi^a(x)},\tilde{X}^R_{{\cal A}^b_\mu(y)}\Bigr] \Bigr] &=& -C_{ab}^c \delta(x-y)\tilde{X}^R_{{\cal A}^c_\mu(x)}
 -m^2C_{ab}^c\lambda_c\delta^\mu_0\delta(x-y)\Xi\nn\\
\Bigl[\Bigl[\tilde{X}^R_{\varphi^a
(x)},\tilde{X}^R_{{\cal E}^b_j(y)}\Bigr] \Bigr] &=& -C_{ab}^c
\delta(x-y)\tilde{X}^R_{{\cal E}^c_j(x)}
-\delta_a^b\partial_j^x\delta(x-y)\Xi\nn\\
\Bigl[\Bigl[\tilde{X}^R_{{\cal A}^a_j(x)},\tilde{X}^R_{{\cal E}^b_k(y)}\Bigr]
\Bigr] &=& -\eta^{jk}\delta_{ab}\delta(x-y)\Xi\,,\label{ym-com}
\eea
where here, in contrast with the Abelian situation, double bracket indicates the commutator in the Lie algebra
$G_{MYM}$, to avoid confusion with the commutator in the Lie algebra of $G$,
and $\eta^{\mu\nu}={\rm diag}(1,-1,-1,-1)$ stands for the
Minkowski metric. In the same way we derive the left-invariant
vector fields:
\bea
\tilde{X}^{L}_{\varphi(x)}&=&X^{L\;(G)}_{\varphi(x)}+\partial^\mu(U^\dag {\cal F}_{\mu\nu}U)n^\nu\Xi\nn\\
\tilde{X}^{L}_{\vartheta_\mu(x)n^\mu}&=&U\frac{\!\!\delta}{\delta \vartheta_\mu(x)n^\mu}U^\dag-m^2(U^\dag\lambda U-\lambda)\Xi\nn\\
\tilde{X}^{L}_{{\cal A}_\mu(x)}&=&U\frac{\!\!\delta}{\delta
{\cal A}_\mu}U^\dag+\left(m(U^\dag\lambda U-\lambda)\eta^{\mu\nu}-
U^\dag {\cal F}^{\nu\mu}U\right)n_\nu\Xi\nn\\
\tilde{X}^{L}_{{\cal F}_{\mu\nu}(x)}&=&U\frac{\!\!\delta}{\delta {\cal F}_{\mu\nu}(x)}U^\dag+\frac{1}{2}[U^\dag({\cal A}^\nu-\vartheta^\nu)Un^\mu-U^\dag({\cal A}^\mu-\vartheta^\mu)Un^\nu]\Xi\nn\\
\tilde {X}^{L}_{\zeta} &=& \hbox{Re}(i\zeta \frac{\partial}{\partial\zeta})\equiv \Xi\,.\label{ym-XL}
\eea

Directly from the group law or by duality on (\ref{ym-XL}) the
left-invariant $1$-form in the $\zeta$-direction,
$\Theta^G_{MYM}$, can be computed:
%
\begin{eqnarray}
  \Theta^G_{MYM}&=&\int_\Sigma d{\sigma}^\nu \hbox{Tr}\Bigl(
{\cal F}_{\nu\mu}\delta {\cal A}^\mu - \delta
{\cal F}_{\nu\sigma}({\cal A}^{\sigma}-\vartheta^{\sigma}) +m^2
(\Lambda-\lambda)\delta({\cal A}_\nu-\vartheta_\nu)\qquad \qquad\nn
\\
&+& {\cal F}_{\nu\mu} [\vartheta^{\mu},-i\delta U U^{\dagger}] +
\partial^{\mu} {\cal F}_{\nu\mu}(-i\delta U U^{\dagger})
\Bigr)+\frac{d\zeta}{i\zeta}\nn
\\
&=& \int_\Sigma d{\sigma}^\nu \hbox{Tr}
(\left({\cal F}_{\mu\nu}\delta({\cal A}^\mu-\vartheta^\mu)-({\cal A}^\mu-\vartheta^\mu)\delta
{\cal F}_{\mu\nu}\right)+
    m^2 (\Lambda-\lambda)\delta({\cal A}_\nu-\vartheta_\nu))+\frac{d\zeta}{i\zeta}\,.    \label{TetaG/MYM}
\end{eqnarray}
and from it the Noether invariants ($I=i_{X^R}\Theta^G_{MYM}$):
\bea
I_{{\cal A}^\mu}&=&\left({\cal F}_{\nu\mu}+m^2(\Lambda-\lambda)\eta_{\nu\mu}\right)n^\nu \nn\\
I_{{\cal F}^{\mu\nu}}&=&({\cal A}_\mu-\vartheta_\mu)n_\nu -({\cal A}_\nu-\vartheta_\nu)n_\mu\nn\\
I_{\varphi}&=&\partial^\nu {\cal F}_{\mu\nu}n^\mu+[{\cal A}^\mu-\vartheta^\mu,\,m^2
\Lambda\eta_{\mu\nu}+ {\cal F}_{\nu\mu}]n^\nu \nn\\
I_{\vartheta^\mu n_\mu}&=&m^2 (\Lambda-\lambda). \label{NoetherI}
\eea
For the particular, though standard, choice of the Cauchy surface $n=(1,0,0,0)$ they acquire the expressions:
\bea
I_{{\cal A}^0}&=&m^2(\Lambda-\lambda)\equiv  -\mathbb{E}_0\nn\\
I_{{\cal A}^i}&=&-{\cal F}_{i0}\equiv-\mathbb{E}_i\nn\\
I_{{\cal F}^{i0}}&=&{\cal A}_i-\vartheta_i\equiv\mathbb{A}_i\nn\\
I_{\varphi}&=& -m^2 [{\cal A}^0-\vartheta^0,\,\Lambda]-[{\cal
A}^i-\vartheta^i,\,{\cal F}_{0i}]+\partial^i{\cal F}_{0i}\equiv
-m^2 \mathbb{A}^0-\vec{\nabla}\cdot\vec{\mathbb{E}}+
   [\mathbb{A}^i,\,\mathbb{E}_i]\equiv-\mathbb{G}\nn\\
   I_{\vartheta^0}&=&m^2(\Lambda-\lambda)=I_{{\cal A}^0}\,, \label{NoetherII} \eea
where we have denoted $\mathbb{A}^0\equiv [({\cal
A}^0-\vartheta^0),\,\Lambda]$, the time component of the massive
vector bosons (those perpendicular to $\lambda$), and
$\mathbb{E}_0\equiv -I_{{\cal A}^0}=-m^2(\Lambda-\lambda)$ in
analogy with $\mathbb{E}_i\equiv -I_{{\cal A}^i}$ although we must
be aware that $\mathbb{E}_0$ is intended to be the conjugate pair
of $\mathbb G$ (see later on Eq. (\ref{MYM-PB})).


We should observe that the Noether invariants $I_{\vartheta^0}$ and
$I_{{\cal A}^0}$ coincide, so that the group parameter ${\cal A}^0-\vartheta^0$ is
actually a gauge parameter in the strict sense (the corresponding
subgroup leaves the solution manifold invariant point-wise).
Notice also that the Noether invariant $I_{{\cal A}_0}$ in the direction
of the $H$ subalgebra is a function of the invariants $I_{{\cal A}_0}$ in
the direction of the $G/H$ orbit (the same dependence occurs for
$I_{\vartheta_0}$). Then, the independent parameters in the solution
manifold are:
\be
(\mathbb{A}^a_i, \mathbb{E}^a_j, \mathbb{E}^b_0, \mathbb{G}^b)\,\;\;a=1,...,\hbox{dim}\;G,\;\;b=1,...,\hbox{co-dim}\;H\,.\label{variables}
\ee
The $1$-form (\ref{TetaG/MYM}) naturally comes down to the quotient $G/H$, which can be
identified with the coadjoint orbit of $G$,
and it can also be written  in terms of Noether invariants. In
fact, we have
\be
\Theta^G_{MYM}=\int_\Sigma d\sigma^\nu\hbox{Tr}\left\{(I_{{\cal A}_\mu}\delta I_{{\cal F}^{\mu\nu}}-
I_{{\cal F}^{\mu\nu}}\delta I_{{\cal A}_\mu})+ I_{\vartheta_\mu n^\mu}\delta([I_\varphi,I_{{\cal A}^\nu}])\right\} +\frac{d\zeta}{i\zeta}\,.
\ee
%

According to the Poisson bracket definition (\ref{chochodef}), the
Noether invariants (\ref{NoetherI}) close a Poisson algebra
isomorphic to the Lie algebra of the symmetry group $G_{MYM}$. The
non-zero brackets are:
\bea \left\{\mathbb{G}_a(\vec{x}),\mathbb{G}_b(\vec{y})\right\}
&=& -C_{ab}^c
\mathbb{G}_c(\vec{x})\delta(\vec{x}-\vec{y}),\nn\\
\left\{\mathbb{A}^j_a(\vec{x}),\mathbb{E}^k_b(\vec{y})\right\} &=&
\eta^{jk}\delta_{ab}\delta(\vec{x}-\vec{y}),\label{MYM-PB}\\
\left\{\mathbb{G}_a(\vec{x}),\mathbb{A}^j_b(\vec{y})\right\} &=&
-C_{ab}^c \mathbb{A}_c^j(\vec{x})\delta(\vec{x}-\vec{y})
+\delta_{ab}\partial^j_x\delta(\vec{x}-\vec{y}),\nn\\
 \left\{\mathbb{G}_a(\vec{x}),\mathbb{E}_b^\mu(\vec{y})\right\} &=&
 -C_{ab}^c \mathbb{E}_c^\mu(\vec{x})\delta(\vec{x}-\vec{y})
 +m^2\delta^\mu_0C_{ab}^c\lambda_c\delta(\vec{x}-\vec{y})\nn\,.
\eea
They reproduce the standard Poisson brackets of the non-Abelian
Yang-Mills theory  \cite{Itzykson,Jackiw} , along with those
corresponding to the new variables $\mathbb E^a_0$, which are
absent from the conventional theory.

 Even though we have not included
space-time translations (nor Lorentz transformations) explicitly
in the group law, this Poisson algebra can be added with a quartic
function, ${\mathbb H}$, of the Noether invariants, constituting
the classical Hamiltonian in our group-theoretical scheme. In
fact, the Hamiltonian function
\be \mathbb{H}=\frac{1}{2}\int
\d^3x\tr\left[(\vec{\mathbb{E}}^2+\vec{\mathbb{B}}^2)+
m^2[\mathbb{A}_0,\,\Lambda]^2+m^2([\vec{\mathbb{A}},\,\Lambda]+
\vec{\nabla}\Lambda)^2-2\mathbb{A}_0\mathbb{G}\right],
\label{Amiltonio} \ee
with $\Lambda=-\frac{\mathbb E_0}{m^2}+\lambda$, actually
reproduces the Lagrangian equations of motion
(\ref{eqofmotSYM-PT}) by making use of the Poisson brackets
(\ref{MYM-PB}) and the correspondence
\[ E_i\leftrightarrow\mathbb{E}_i,\;\;\Lambda\leftrightarrow\Lambda,\;\;
A_i\leftrightarrow \mathbb{A}_i,\;\; A_0-\theta_0\leftrightarrow
\mathbb{A}_0\]
and remembering that $[\vartheta_\mu,\,\Lambda]=\partial_\mu\Lambda$.
In fact:
\begin{align} \dot{\mathbb{A}}^i(x)&=\{\mathbb{A}^i(x),\mathbb{H} \}=-\mathbb
E^i(x)+\partial^i\mathbb{A}^0(x)+[\mathbb{A}^0,\mathbb A^i](x)\nn\\
 \dot{\mathbb{E}}^i(x)&=\{\mathbb{E}^i(x),\mathbb{H} \}=
 \epsilon^{ij}_{\phantom{ij}k}\partial_j
 \mathbb B^k(x)+\epsilon^{ij}_{\phantom{ij}k}[\mathbb A_j,\mathbb
 B^k](x)+[\mathbb{A}^0,\mathbb
 E^i](x)+m^2[\partial^i\Lambda-[\mathbb A^i,\Lambda],\Lambda](x)\nn\\
 \dot \Lambda(x)&=\{\Lambda(x),\mathbb{H} \}=\frac{1}{m^2}[\mathbb
 G,\Lambda](x)\\
  \dot{\mathbb G}(x)&=\{\mathbb G(x),\mathbb{H} \}=0.\nn
\end{align}
To be precise, the first equation just constitutes the definition
of the electric field. The second one literally corresponds to the
$\nu=i$ component of (\ref{eqofmotSYM-PT}), whereas the third one
is the projection on the massive internal components (expressed
 by means of a commutation with $\Lambda$) of the $\nu=0$ index of
  (\ref{eqofmotSYM-PT}) (i.e., the Gauss law). Note that the Gauss Law in the internal massless direction 
(of the $H$ subgroup) is neither a Hamiltonian equation in the standard variational formulation. Finally, the last
  equation just states the gauge invariance of our Hamiltonian. 


\subsubsection{The Quantum Representation}\label{quantumsubsecSU(2)}

According to the general scheme of GAQ we start from complex
$U(1)$-functions $\Psi$ on the entire centrally-extended group to
be represented, that is, functions which are homogeneus of degree
one on the parametre $\zeta\in U(1)$. In order to obtain an
irreducible representation these functions must be restricted by
the Polarization condition established by means of a polarization
subgroup $G_P$ of the finite left action. A look at the Lie
algebra commutators of our symmetry group reveals that the actual
Characteristic subgroup is constituted by the following elements:
\be g_{{\cal C}}=(U_H,\vartheta_\mu n^\mu,{\cal
A}_\mu=\vartheta_\mu,{\cal F}_{\nu\sigma}=0,\zeta=1)\,, \ee
whereas the Polarization subgroup is constituted by the following
elements:
\be g_P=(U_H,\vartheta^\mu n_\mu, {\cal A},{\cal F}=0,\zeta=1)\,,
\ee
which act on the original complex functions $\Psi(g')=\Psi(U',\vartheta'_\mu n^\mu,{\cal A}'_\nu, {\cal F}'_{\mu\nu},\zeta')$
from the right: $\Psi(g')\rightarrow\Psi(g'g_P)$.

The key point in searching for the appropriate form of the wave
functions, invariant under the Polarization subgroup, is to notice
the factor which appears in front of the wave function as a
consequence of the cocycle in the composition law of the $U(1)$
argument ($\Psi$ is homogeneous of degree one on it). This factor
must be cancelled  out by some $U(1)$ term factorizing an
arbitrary function $\Phi$ of given arguments. It is easy to see
that the following  wave function $\Psi$ is the general solution
to the polarization equation (\ref{PolarizacionF}):
\be \Psi=\zeta e^{i\int_\Sigma d\sigma^\mu\hbox{Tr}\left(m
\lambda(U^\dag\check{{\cal A}}_\mu U-\check{{\cal A}}_\mu)+
      \frac{1}{2}\check{{\cal A}}^\nu {\cal F}_{\mu\nu}\right)}
      \Phi((\Lambda-\lambda),{\cal F})\label{wavefunction(su2)}
\ee
where $\check{{\cal A}}\equiv({\cal A}-\vartheta)$ and $\Phi$ is
an arbitrary function of its arguments. In fact, choosing
$d\sigma^\mu$ as $d^3x$, without loss of generality,
\bea \Psi(g'g_P)&=&\zeta'e^{i\int d^3x\hbox{Tr}
\left[m\lambda\left((U'U_H)^\dag(U'\check{{\cal
A}}_0U'^\dag+\check{{\cal A}}_0')(U'U_H)-U'\check{{\cal
A}}_0U'^\dag-\check{{\cal A}}_0'\right)
+\frac{1}{2}(U'\check{{\cal A}}^\nu U'^\dag+\check{{\cal A}}'^\nu){\cal F}'_{0\nu}\right]}\nn\\
&\times& e^{i\int d^3x\hbox{Tr}\left[m\lambda(U'\check{{\cal A}}_0U'^\dag-\check{{\cal A}}_0)-
\frac{1}{2}{\cal F}'_{0\nu}U'\check{{\cal A}}^\nu U'^\dag\right]}
\Phi(U'U_H \lambda(U'U_H)^\dag-U_H\lambda U_H^\dag,\,{\cal F}')\nn\\
&=&\Psi(g') \eea
%

%
For the usual choice of $\Sigma$, the arbitrary part of $\Psi$ can
be written in terms of the variables (\ref{variables}).
$\Phi=\Phi(\mathbb{E}^a_\mu)$, if we adopt the convention that
$\mathbb{E}^c_0=0\,,\;\;\forall c$ running on $H$. That is, we
arrive at the (generalized) ``electric field representation''.

The action of the right-invariant vector fields preserve the space of polarized wave functions, due to the commutativity of the left and right actions
as  already stated, so that it is possible to define an action of them on the arbitrary factor $\Phi$ in the wave functions. It is not difficult to
demonstrate that on this space of functions the quantum operators acquire the following expression:
\begin{align}
\hat{E}^\mu_a\Phi&\equiv\phantom{-} i \zeta^{-1} e^{-i\int_\Sigma
d\sigma^\mu \hbox{Tr}\left(m^2\lambda(U^\dag\check{{\cal A}}_\mu
U-\check{{\cal A}}_\mu)+\frac{1}{2}\check{{\cal A}}^\nu {\cal
F}_{\mu\nu}\right)}\tilde{X}^R_{{\cal A}^a_\mu}\Psi
=\mathbb{E}^\mu_a\Phi\nn\\
\hat{A}^i_a\Phi&\equiv -i \zeta^{-1}e^{-i\int_\Sigma d\sigma^\mu
\hbox{Tr}\left(m^2\lambda(U^\dag\check{{\cal A}}_\mu
U-\check{{\cal A}}_\mu)+\frac{1}{2}\check{{\cal A}}^\nu {\cal
F}_{\mu\nu}\right)}\tilde{X}^R_{{\cal F}^a_{0i}}\Psi
=-i\frac{\!\!\delta}{\delta \mathbb{E}^i_a}\Phi\label{operatas(su2)}\\
\hat{G}_a\Phi&\equiv -i \zeta^{-1} e^{-i\int_\Sigma d\sigma^\mu
\hbox{Tr}\left(m^2\lambda(U^\dag\check{{\cal A}}_\mu
U-\check{{\cal A}}_\mu)+\frac{1}{2}\check{{\cal A}}^\nu {\cal
F}_{\mu\nu}\right)}\tilde{X}^R_{\varphi^a}\Psi\nn\\
&=\left(\vec{\nabla}\cdot\vec{\mathbb{E}}_a +
C^c_{ab}\left((\mathbb{E}^b_0-m^2\lambda^b)
\frac{\!\!\delta}{\delta \mathbb{E}^c_0}-
\vec{\mathbb{E}}^b\cdot\frac{\!\!\delta}{\delta\vec{\mathbb{E}}^c}\right)\right)\Phi\nn\\
\hat{A}^0\Phi &\equiv
\frac{1}{m^2}(\hat{G}-\vec{\nabla}\cdot\hat{\vec{{E}}}+[\hat{A}^j,\,\hat{E}_j])\Phi\nn
\end{align}
The constraint condition $\hat{G}\Phi=0$, if required, would
account for the quantum implementation of the non-Abelian ``Gauss
law".

On the quantum representation space we can construct the Hamiltonian operator $\hat{H}$ that
represents, without ambiguity, the classical Hamiltonian (\ref{Amiltonio}):
\be \hat{H}=\frac{1}{2}\int
\d^3x\tr\left(\hat{\vec{E}}^2+\hat{\vec{B}}^2+m^2[\hat{A}_0,\frac{\hat{E}_0}{m^2}-\lambda]^2
+m^2(\frac{\vec\nabla\hat{E}_0}{m^2}+[\hat{\vec
A},\,\frac{\hat{E}_0}{m^2}-\lambda])^2-2\hat{A}_0\hat{G}\right)\,.\label{QAmiltonio}
\ee
It is of remarkable relevance the fact that this operator
preserves the Hilbert space of quantum states.

It should be stressed that the central term proportional to
$\lambda_c$ in the last bracket  of (\ref{MYM-PB}) could also be
considered as a remnant of some sort of ``symmetry breaking'' in
the sense that it can be hidden into a redefinition of
$\hat{E}_a^0$,
\be \hat{E}_a^0\rightarrow
\hat{E}'^0_a\equiv\hat{E}_a^0-m^2\lambda_a=-m^2\Lambda_a,\label{shiftE} \ee
which now acquires a non-zero vacuum expectation value
proportional to the mass $m^2\lambda_a$, that is:
\be\langle 0|\hat{E}_a^0|0\rangle=0\longrightarrow \langle
0|\hat{E}'^0_a|0\rangle=- m^2\lambda_a.\ee

\section{Connection with more standard techniques and concluding remarks}\label{Perturbation}

In the present paper we have provided a consistent quantization of
the Poisson algebra among the basic functions on the solution
manifold of massive Yang-Mills fields coupled to non-linear
partial-trace sigma scalar fields. This quantization has been
achieved through a group-quantization approach which looks for
unitary and irreducible representations of a proper symmetry
group, the group named $G_{MYM}$ in the text. As already
commented, the parameters of this group have been written as
functions on the Cauchy hyper-surface $\Sigma$,  and we may call
${\cal H}_\Sigma$ the Hilbert space made of (square integrable)
functions of the arguments of the wave functions.  In the present
case they are ``electric fields'' $\mathbb E$ with arguments on
$\Sigma$, although we commented that they could have equally be
chosen as functions on the entire Minkoswski space-time subjected
to their (classical) equations of motion. In fact, a quantum
Hamiltonian operator $\hat{H}$, uniquely defined in terms of the
basic operators, which preserves the quantum representation space,
that is, the Hilbert space ${\cal H}_\Sigma$, has been
constructed. This allows, in principle, any proper computation
concerning the time evolution.

To be more specific and, in order to connect our scheme with more
standard settings, let us translate our Hilbert space to an
isomorphic Hilbert space made of wave functions with arguments
defined on the entire Minkowski space-time but, that time,
satisfying the Schr\"odinger equation associated with our
Hamiltonian operator $\hat{H}$ given in (\ref{QAmiltonio}),
\[ i\hbar\frac{\partial\Phi}{\partial t}=\hat{H}\Phi\,. \]
The space of solutions of the Schr\"odinger equations with initial
conditions in ${\cal H}_\Sigma$, to be named simply ${\cal H}$,
though equivalent, proves to be more appropriate to establish any
comparison with standard computations. It is formally constructed
by means of the unitary lifting-in-time operator $U(t)$ defined as
$U(t)\equiv \exp({-i\hbar t\hat{H}})$, so that wave functions and
quantum operators are lifted from ${\cal H}_\Sigma$ to ${\cal H}$
with the traditional expressions
\[ \Psi(t)=U(t)\Psi,\; \;\hat{O}(t)=U(t)\hat{O}U^\dag(t)\,. \]
This way, we are in conditions to write down an exact formal
expression for the arbitrary-time commutators between basic field
operators $\hat A^\mu_a(\vec{x},t)$:
\[[\hat A^\mu_a(\vec{x},t),\,\hat A^\nu_b(\vec{y},t')]=
[U(t)\hat A^\mu_a(\vec{x})U^\dag(t),\,U(t')\hat
A^\nu_b(\vec{y})U^\dag(t')]\]
and analogously for $\hat E^\mu_a$ or any other. Also, a unitary
evolution operator, in the usual way, acting on ${\cal H}$ is
trivially constructed by the expression $U(t,\,t')\equiv U^\dag(t)
U(t')$, that which permits the formulation of a $S$ matrix.

As far as the explicit way of computing $U(t)$ and $U(t,\,t')$ is
concerned, we must remark that the operator $\hat{H}$, itself, is
well-behaved on the representation space of our original symmetry
group $G_{MYM}$, but some part of it, $\hat{H}_{int}$, intended to
describe ``the interaction'' might be ill-defined (this comment
also applies to standard perturbation theory, \`a la Dyson, in
dealing with typically non-linear theories like, precisely, those
involving non-linear sigma Lagrangian terms). Therefore we are
aimed at working from a perturbative framework which uses the
entire Hamiltonian as developed, for instance, in traditional
textbooks collections like \cite{Landau}. There, however, the
exact propagator defined in terms of basic (exact) operators as
\[ {\cal D}(x,\,x')=\langle 0|T\hat A(x)\hat A(x')|0\rangle \]
is further developed in series of the free propagator $D(x,x')$
thus connecting with the Interaction-Image series associated with
standard Dyson series. Unfortunately, this expansion in terms of
free field objects, like free propagator, cannot be properly
achieved mainly due to the ``sigma sector'' related to the fields
with index $0$ in the theory, that is, $\hat{E}^0$ and the
conjugate operator $\hat{G}$, for which the commutation relations
are non-canonical. In fact, in the standard perturbative framework
the Wick theorem is widely used in providing  the Feymman rules,
but it requires that creation and annihilation operators undergo
canonical commutation relations. In fact, if this is not the case,
the Wick contraction does not reduce the number of operators and
an infinite chain of additional terms appear in the computation of
the $S$ matrix. In addition, the standard perturbative expansion
departs from ``in'' states which are intended to verify a
Klein-Gordon-like equation but the equation of motion for the
``unperturbed'' sigma fields admits an extra infinite set of
solutions which do not correspond to massless Klein-Gordon fields.
In this respect see Ref. \cite{Pertursigma}.

Therefore, to compute operations involving exact objects we must resort to another more appropriate technique, more directly related to the
group-theoretical scheme here developed. To be precise, the Magnus expansion \cite{Magnus} of the operator $U(t)$ provides a perturbative
scheme which preserves unitarity at each order in the exponential of the Hamiltonian. In this type of perturbation, each order of approximation
is dictated by the powers in the enveloping algebra of the original one, rather than powers in a supposedly small interaction constant. The
general form of this expansion is as follows:
\be
  \hat U(t) = e^{\hat \Omega (t)}\,, \qquad \hat \Omega(0) = 0 \, ,
\ee
\begin{align}
   \hat \Omega (t) &= \lim_{n\to\infty}\hat \Omega^{[n]}(t)\\
\hat \Omega^{[n]} (t) &= \sum_{k=0}^{\infty}\frac{B_k}{k!}\int^{t}_{0} \d t_1
   {\rm ad}^k_{\hat \Omega^{[n-1]}(t_1)}
       \bigl(-\frac{i}{\hbar}\hat H(t_1)\bigr)
\label{eq:Omegan}\,,
\end{align}
where $B_k$ are the Bernoulli numbers and
\begin{equation}
{\rm ad}^0_{\hat A} (\hat B)\equiv \hat B\,, \qquad {\rm
ad}^1_{\hat A} (\hat B)\equiv [\hat A, \hat B]\,, \qquad {\rm
ad}^k_{\hat A}(\hat B)\equiv [{\rm ad}^{k-1}_{\hat A}(\hat
B),\,\hat B]\,.
\end{equation}
However, in our particular case in which the Hamiltonian is independent of time, the expression for the time-lifted version of
an operator $\hat{O}(t)$ acquires the simpler form:
\begin{equation}
  \hat{O}(t) = e^{\hat{\Omega}(t)}\hat{O}e^{-\hat{\Omega}(t)} =
    \sum_{k=0}^{\infty} \frac{1}{k!}(\frac{-it}{\hbar})^k
     {\rm ad}^k_{\hat{H}} \bigl(\hat{O} \bigr)\,,
\end{equation}
expression which can be further reduced depending on the way in which the Hamiltonian closes algebra with the operators $\hat{O}$.
This extent is soundly analyzed in Ref. \cite{Wilcox}.

It is remarkable that the Magnus expansion approach naturally
suggests a more algebraic procedure which has been indeed used in
the actual computation of physical quantities appearing in
conformal field theories bearing high symmetries like string
theories or Wess-Zumino-Witten systems \cite{Kac-Moody-like}. In
fact, starting from our algebra of basic operators
(\ref{operatas(su2)}), we may proceed by closing a new Lie algebra
by commutation with $\hat{H}$, order by order in a formal
``expansion'' constant $\alpha$, exponentiating the resulting
algebra up to the same order in $\alpha$, and re-quantizing again
with the present group-theoretical method. Much work is being made
in this direction by the authors with the present symmetry
\cite{inpreparation}.

Let us mention our group-theoretical viewpoint as regard another
algorithm usually assumed as basic in the quantization of
non-Abelian gauge fields. We refer to the use of BRST symmetry to
address the constraints, which is absent from our present scheme.
Firstly, the employ of ghost and anti-ghost field is somehow
motivated by a more simple way of rewriting the measure on the
space of classical trajectories in the Feymman path approach to
quantum theory of gauge fields. On the other hand, however, much
work on BRST super-symmetry  has been developed by some of the
authors and collaborators, concluding that this extra symmetry
seems not to play an essential physical role. Really, in Ref.
\cite{ElectroBRST}, a Lagrangian for the Abelian gauge theory,
including a covariant gauge fixing term as well as a ghost term,
from a group law incorporating the BRST symmetry was derived
according to the present formalism. Then, in Ref.
\cite{NonAbelianBRST} a supergroup law was found, including the
BRST fermionic parameter as well as fermionic ghosts and
anti-ghosts, for an arbitrary semi-simple group of constraints,
the Lie super-algebra of which generalizes the algebra of Bowick
and Gursey \cite{Bowick}. Also, in Ref. \cite{JMPBRST}, this sort
of supergroups were applied to some systems, including Virasoro
constraints, and it was demonstrated that the quantization of such
systems can be achieved equivalently either with BRST machinery or
without it; the BRST-cohomology seemed to be more a question of
fashion, let us say, rather than a real necessity.

Just to finish, we remark that the present treatment of massive non-Abelian gauge theory,
in either perturbative schemes mentioned above, would be specially suited to achieve a 
consistent approach to the quantum description of the Physics around the Standard Model 
of Electro-Weak interactions without the Higgs particle, and that this is
the final target of our present paper. It should be stressed that the two 
examples here analyzed (Abelian $U(1)$  and non-Abelian partial-trace
$SU(2)$ cases) accomplish this task provided that the gauge field 
associated with hyper-charge acquires mass according to the scheme developed in 
Sec. \ref{quantumsecU(1)}   and the Lie algebra (rigid) generator $\lambda$ in Sec. \ref{quantumsecSU(2)} 
is chosen in the mixed ``electric-charge'' direction in the Lie algebra of $SU(2)\times U(1)$, as explained in Sec. 4 of Ref. \cite{SigmaYM}.

\section*{Acknowledgements}

 Work partially supported by the
Spanish MICINN, Fundaci\'on S\'eneca and Junta de Andaluc\'\i a
under projects  FIS2008-06078-C03-01, 08814/PI/08 and
FQM219-FQM1951, respectively.


\end{document}